\documentclass[journal]{IEEEtran}
\usepackage{graphicx,amsmath,amsfonts,amsthm,amssymb,cite,multirow,multicol,booktabs}

\begin{document}

\title{Privacy-Preserving Brain-Computer Interfaces:\\ A Systematic Review}

\author{Kun~Xia, Wlodzislaw Duch, Yu~Sun, Kedi~Xu, Weili~Fang, Hanbin~Luo, Yi Zhang, Dong Sang,\\ Xiaodong Xu$^*$, Fei-Yue Wang$^*$ and Dongrui~Wu$^*$

\thanks{K.~Xia~and~D.~Wu are with the Ministry of Education Key Laboratory of Image Processing and Intelligent Control, School of Artificial Intelligence and Automation, Huazhong University of Science and Technology, Wuhan 430074, China. D.~Wu is also with Zhejiang Lab, Hangzhou 311121, China. Email: kxia@hust.edu.cn, drwu@hust.edu.cn.}
\thanks{W. Duch is with the Department of Informatics, School of Physics, Astronomy and Informatics, and Neurocognitive Laboratory, Center for Modern Interdisciplinary Technologies, Nicolaus Copernicus University, Torun, Polan. Email: wduch@umk.pl.}
\thanks{Y.~Sun is with the Ministry of Education Key Laboratory for Biomedical Engineering, Zhejiang University, Hangzhou, China. He is also with the Zhejiang University School of Medicine, National Clinical Research Center for Child Health, Hangzhou, China. Email: yusun@zju.edu.cn.}
\thanks{K.~Xu is with Zhejiang Lab, Hangzhou 311121, China. He is also with Qiushi Academy for Advanced Studies (QAAS), Department of Biomedical Engineering, Zhejiang University, Hangzhou, China. Email: xukd@zju.edu.cn.}
\thanks{W.~Fang and H.~Luo are with the School of Civil and Hydraulic Engineering, Huazhong University of Science and Technology, Wuhan 430074 China. Email: weili\_f@hust.edu.cn, luohbcem@hust.edu.cn.}
\thanks{Y.~Zhang and X.~Xu are with the College of Public Administration, Huazhong University of Science and Technology, Wuhan, 430074 China. Email: yizhanghn@sina.com, xiaodong-xu@hust.edu.cn.}
\thanks{D. Sang is with the NetsUnion Clearing Corporation, Beijing 100045, China. Email: sd1984r@163.com.}
\thanks{F-Y Wang is with the State Key Laboratory for Management and Control of Complex Systems, Institute of Automation, Chinese Academy of Sciences, Beijing 100049, China. Email: feiyue.wang@ia.ac.cn.}
\thanks{X. Xu, F-Y Wang and D.~Wu are the corresponding authors.}}

\maketitle

\begin{abstract}
A brain-computer interface (BCI) establishes a direct communication pathway between the human brain and a computer. It has been widely used in medical diagnosis, rehabilitation, education, entertainment, etc. Most research so far focuses on making BCIs more accurate and reliable, but much less attention has been paid to their privacy. Developing a commercial BCI system usually requires close collaborations among multiple organizations, e.g., hospitals, universities, and/or companies. Input data in BCIs, e.g., electroencephalogram (EEG), contain rich privacy information, and the developed machine learning model is usually proprietary. Data and model transmission among different parties may incur significant privacy threats, and hence privacy protection in BCIs must be considered. Unfortunately, there does not exist any contemporary and comprehensive review on privacy-preserving BCIs. This paper fills this gap, by describing potential privacy threats and protection strategies in BCIs. It also points out several challenges and future research directions in developing privacy-preserving BCIs.
\end{abstract}

\begin{IEEEkeywords}
Brain-computer interfaces, privacy, machine learning, electroencephalogram
\end{IEEEkeywords}

\IEEEpeerreviewmaketitle

\section{Introduction}

A brain-computer interface (BCI) constructs a directly communication pathway between the human brain and a computer. It has been used in clinical diagnosis, rehabilitation, fatigue monitoring, gaming, robot control, etc~\cite{Gu2020}. For example, a BCI system can be used to estimate seizure onset time and location, so that appropriate drugs or deep brain stimulation can be delivered~\cite{Li2020a}. Cochlear implants~\cite{Zeng2008}, one of the most successful invasive BCI systems for restoring nerve functions (as of 2020, about 800,000 people worldwide had received such implants), directly stimulate the cochlear nerve through electrical impulses to transmit auditory information to the brain. BCIs have also been used to estimate the relevance of information during reading and thus may create a profile of user interest \cite{Eugster2016}, and to monitor the user's mental states for intelligent assistive systems, which are useful in education and healthcare~\cite{Ienca2017}.

The effectiveness of BCIs relies on the rich information contained in input signals, which contain private and sensitive medical and personal information~\cite{Neupane2019}. As a result, there are more and more privacy attacks targeting BCIs. For example, Palazzo \emph{et al.}~\cite{Palazzo2017} showed that a user's brain electrical signals can be used to reconstruct the scene he/she saw. What intensifies privacy threat is that developing and using a BCI system may require data transmission among different users, developers and organizations. For example, a hospital may share some electroencephalogram (EEG) databases with a university or company to help develop a new medical system. However, by directly sharing EEG data with others, some malicious attackers may learn sensitive information irrelevant to the main BCI task, including personality traits and cognitive abilities~\cite{Frank2017,Bellman2018,Landau2020}, leading to the leaking of private patient information.

Although most BCI research so far still focuses on its accuracy and efficiency, the privacy of BCIs started to attract attentions gradually. In 2015, Li \emph{et al.}~\cite{QianQianLi2015} analyzed potential privacy threats in four BCI application scenarios, i.e., neuromedical applications, user authentication, gaming and entertainment, and smartphone-based applications, and introduced some counter-measures. In 2016, Takabi \emph{et al.}~\cite{Takabi2016} analyzed privacy issues in BCI App stores, including software development kits (SDKs), application programming interfaces (APIs), and BCI applications. They found that ``\emph{most applications have unrestricted access to users' brainwave signals and can easily extract private information about their users without them even noticing.}" In 2017, Yuste \textit{et al.} \cite{Yuste2017} commented on four ethical priorities for neurotechnologies and artificial intelligence in \emph{Nature}, and explicitly pointed out that ``\emph{artificial intelligence and brain-computer interfaces must respect and preserve people's privacy, identity, agency and equality}." In 2018, Icena \textit{et al.} \cite{Ienca2018} published a commentary in \emph{Nature Biotechnology}, pointing out that direct-to-consumer neurotechnologies, including neuromonitoring headsets (e.g., InteraXon Muse, Emotiv InSight and Epoc+, Neurosky MindWave Mobile 2, and Open BCI Ultracortex Mark IV) and neuromodulation tools (e.g., Foc.Us GoFlow and Focus V3, and Roxon Cefaly), may all have ethical concerns including privacy. On November 1, 2021, the Personal Information Protection Law of China became effective, which explicitly states that all personal information processing parties must strictly protect personal sensitive information, including biometrics and healthcare (EEG falls into these categories). However, to our knowledge, there does not exist a systematic and contemporary review on privacy-preserving BCIs.

This paper fills the above gap. The remainder of it is organized as follows: Section~\ref{sub:information} describes inferrable private information in BCIs. Section~\ref{sub:threats} introduces potential privacy threats in BCIs. Section~\ref{sub:protection} reviews privacy protection approaches in the literature both generic and specific to BCIs. Section~\ref{sub:challenge} points out several challenges in implementing privacy-preserving BCIs. Finally, Section~\ref{sub:conclusion} draws conclusions and points out future research directions.

\section{Private Information in BCIs} \label{sub:information}

\begin{figure}[htpb]   \centering
  \includegraphics[width=.8\linewidth,clip]{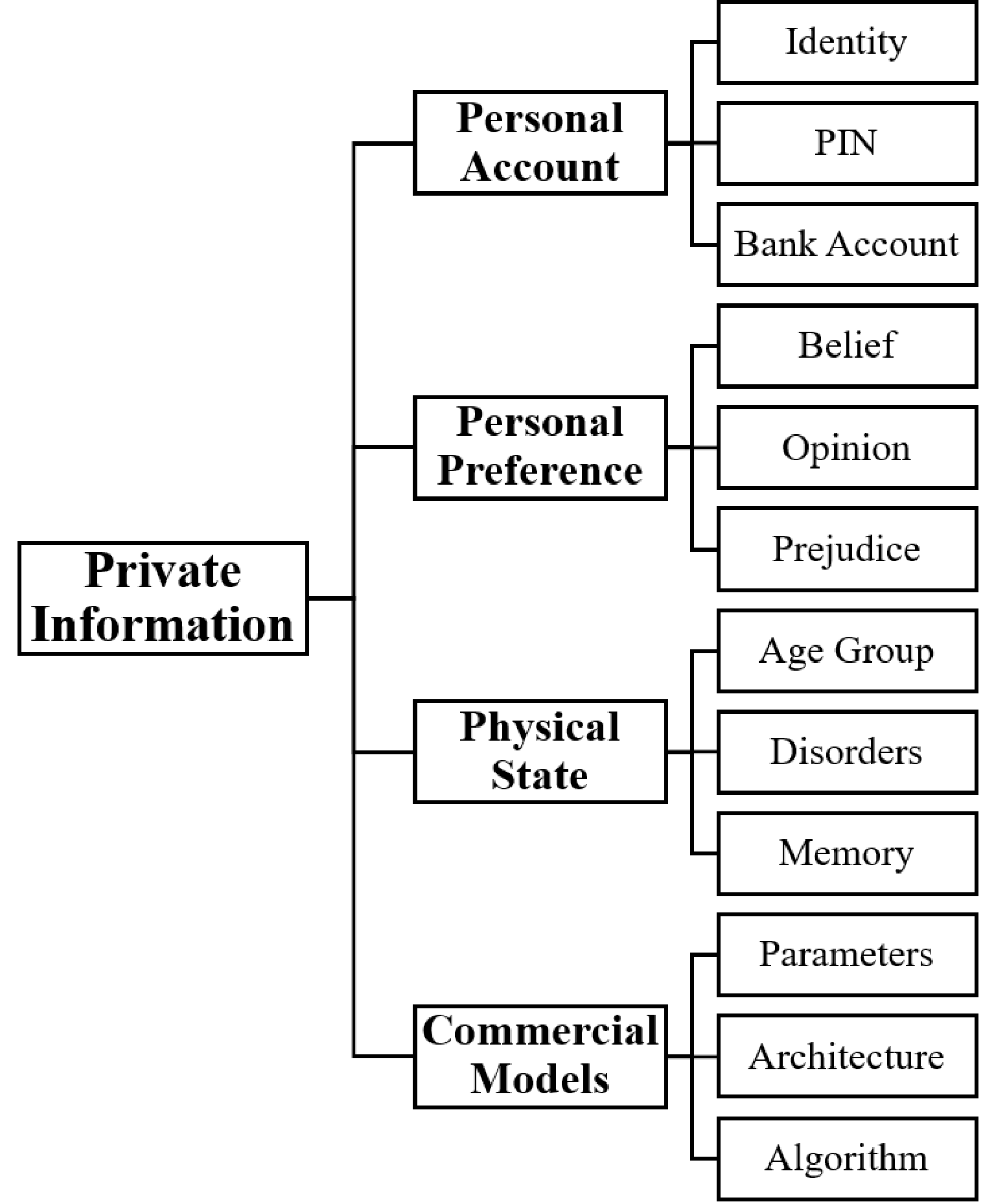}
  \caption{Private information in BCIs.}   \label{fig:PrivateInformation}
  \end{figure}

Private information generally refers to the information that should not be disclosed to others without the permission of the information owner. This section introduces inferrable private information in BCIs, which usually includes four types: personal account, personal preferences, physical state, and commercial models, as shown in Fig.~\ref{fig:PrivateInformation}.

\subsection{Personal Account} \label{subsec:account}

Personal account information may include the user's identity, personal identification number (PIN), bank account, and so on. Therefore, EEG signals can be used for biometric identification \cite{DelPozoBanos2015, RuizBlondet2014}.

Martinovic \emph{et al.}~\cite{Martinovic2012} showed that EEG signals collected by cheap consumer-grade BCI devices can be used to infer private information of users, such as credit cards, PIN numbers, known people, or residential addresses. It is based on the detection of event-related potentials (ERPs), e.g. when a user sees a password that is being used, the EEG signal usually contains a higher ERP.

Neupane \emph{et al.}~\cite{Neupane2017} also used ERPs to eavesdrop a user's numeric PINs or textual passwords. They used two different types of EEG headsets, a consumer-grade device and a higher-end medical device, and three kinds of keyboards, namely a physical keyboard, a virtual keyboard, and an automated teller machine (ATM) style numeric keypad. Their experiments showed that the accuracy is significantly higher than random guessing.

Choi \emph{et al.}~\cite{Choi2018} used resting-state EEG signals to infer the identity of the user. Their experiment showed that this kind of identity authentication can reach an accuracy of 88.4\%. Different from the traditional EEG-based identity authentication system, this authentication does not require external stimuli, that is, it does not require the user's attention. They extracted the changes of the EEG signals in the alpha frequency band ($8-13$Hz) during the opening and closing of the eyes as features, and then calculated the intra- and inter-user cross-correlation to identify users.

\subsection{Personal Preferences} \label{subsec:preference}

Personal preferences usually include beliefs, opinions, prejudices, and so on, which may often be information that one is unwilling to share with others.

Knuston \emph{et al.}~\cite{Knutson2007} found that the activation of amygdala measured by functional magnetic resonance imaging (fMRI) is correlated with the result of Implicit-Association Test about a user's race and gender bias.

Vecchiato \emph{et al.}~\cite{Vecchiato2010} used EEG signals to measure changes in users' brain activities towards different politicians, which can be used to reflect different attitudes towards them. Specifically, when users see images and videos of real politicians, they need to rank the dominance and trustworthiness of the politicians and vote for them. The results showed that users' EEG signals varied in the alpha, beta and gamma frequency bands, so that their political preferences can be inferred. The field of political neuroscience is growing \cite{Jost2014, Schreiber2017}, and BCIs were already used to predict political attitudes and voting behavior in 2019 European Parliament election \cite{Galli2021}.

In addition, Filippi \emph{et al.}~\cite{Filippi2010} showed that EEG signals can be used to analyze the user's emotion for a specific stimulus to infer whether the user has a certain preference or prejudice, e.g., whether the user is a vegetarian.

More recently, Spape \emph{et al.}~\cite{Spape2021} proposed generative BCIs, which integrate generative adversarial networks (GANs) and BCIs to create facial images according to personal preferences. Davis \emph{et al.}~\cite{Davis2021} used collaborative filtering to infer the user's preference on human facial attractiveness from EEGs. They demonstrated that \emph{``preferences can be predicted from EEG/ERP responses in a single-trial setting."}

\subsection{Physical State} \label{subsec:physical}

Physical state mainly includes disorders, age group, smoking status, memory, and so on.

Groves \emph{et al.}~\cite{Groves2017} showed that ERPs detected from the EEG signals when the user is viewing a picture of the body or house are related to the scores on Eating Disorder Inventory-\uppercase\expandafter{\romannumeral2} (EDI-2). Faster visual encoding of both body and control stimuli can be used to infer whether the user has an eating disorder. Neupane \emph{et al. }~\cite{Neupane2019} used EEG signals from users when watching simple pictures and videos to infer whether users had alcohol usage disorder and their age groups with high accuracy. Lutz \textit{et al.} \cite{Lutz2021} pointed out that ERP measures of error processing are candidate biomarkers for disorders that lead to a display of unwanted behaviors, e.g., various addictions.

Similarly, Nestor \emph{et al.}~\cite{Nestor2018} showed that smokers, ex-smokers, and non-smokers have significantly different brain activation patterns in different areas of the brain, which can be measured by fMRI, i.e., brain fMRI can be used to infer a user's smoking status.

\subsection{Commercial Models}  \label{subsec:model}

Private information of a proprietary machine learning (ML) model in a commercial BCI system usually includes its type, structure, parameters, and so on.

BCI system manufacturers generally do not want to leak their ML models, which may expose their systems to white-box adversarial attacks~\cite{Zhang2019a} and affect their market competitiveness. At present, there are no privacy attacks against commercial models in BCIs, but various approaches to stealing models have been developed in other fields~\cite{Tramer2016,CorreiaSilva2018,Papernot2017}. We conjecture that such attacks are also possible for BCI models.

\section{Privacy Threats in BCIs} \label{sub:threats}

This section introduces a typical BCI system and the associated privacy threats.

\subsection{Overview} \label{subsec:overview}

\begin{figure}[htpb]   \centering
\includegraphics[width=\linewidth,clip]{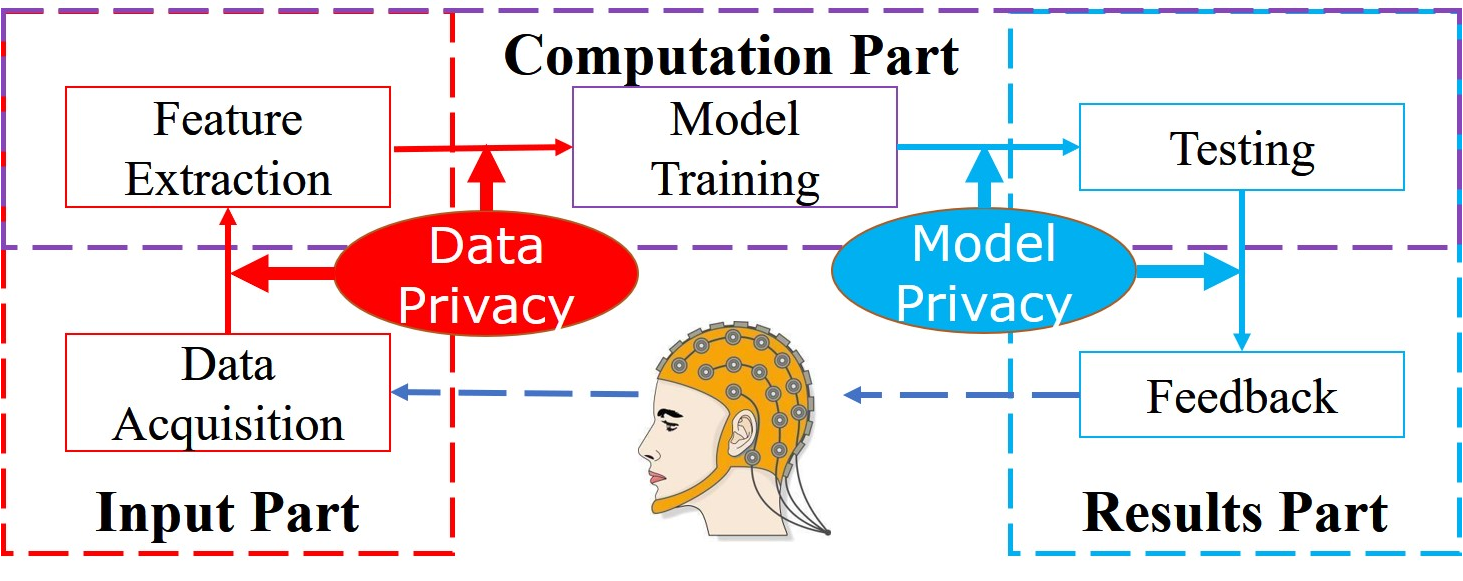}
\caption{Flowchart of a typical BCI system, with corresponding parts and privacy threats.}   \label{fig:PrivacyThreats}
\end{figure}

Similar to an ML system~\cite{AlRubaie2019}, a typical BCI system consists of three parts, as shown in Fig.~\ref{fig:PrivacyThreats}: the input part (data owners or users), the computation part, and the results part.

The input part often performs data acquisition and feature extraction, where feature extraction may be conducted by the computation part. The computation part is responsible for model training. The results part uses its data for testing. A closed-loop BCI system includes a feedback path to the user, and the results part and the input part are the same.

\begin{figure}[htpb]   \centering
  \includegraphics[width=.8\linewidth,clip]{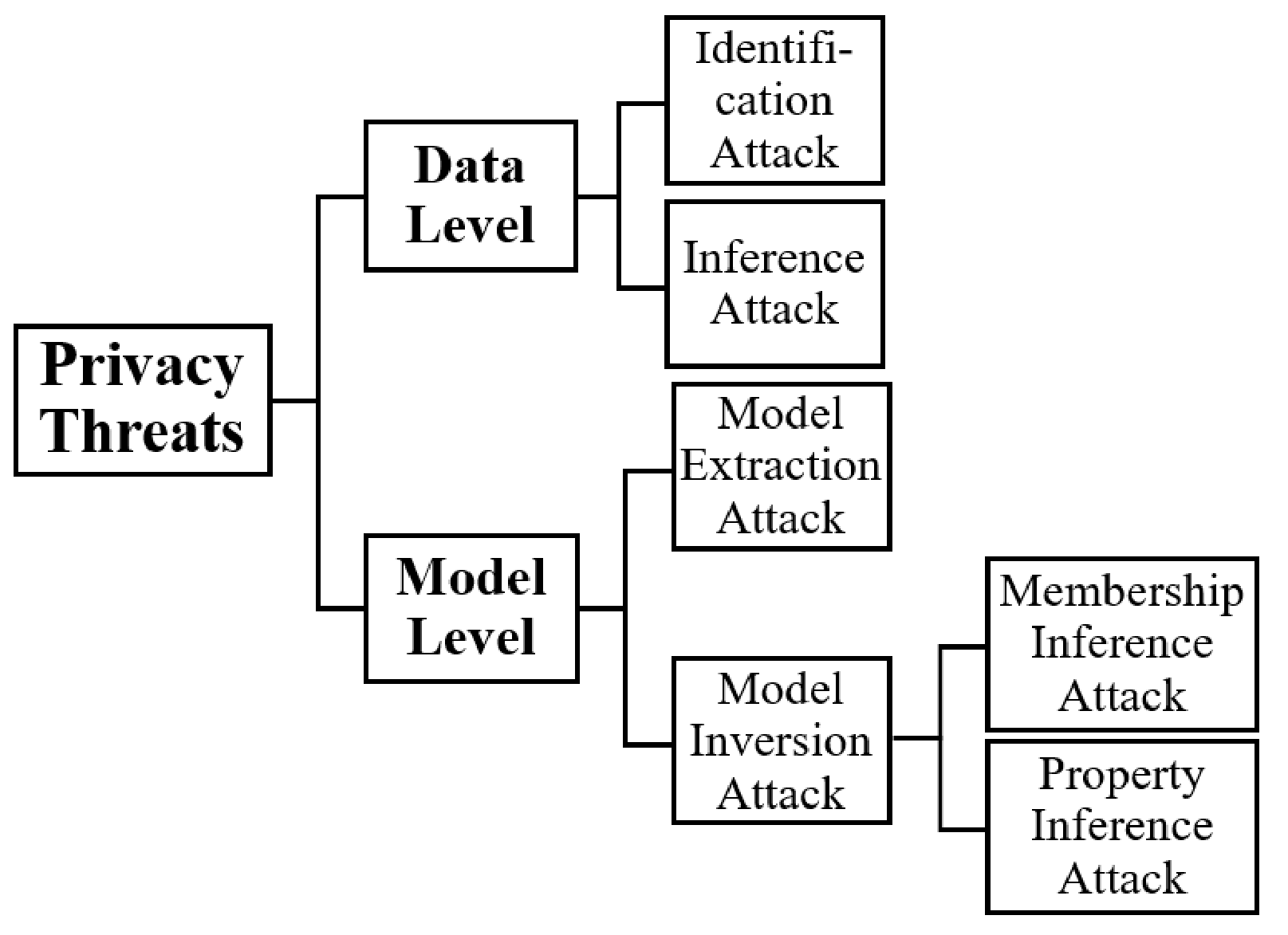}
  \caption{Privacy threat types.}   \label{fig:PrivacyThreatType}
  \end{figure}

Privacy threats may occur during data transmission from the input part to the computation part, or during model delivery in the test phase. According to the attack target, there are two types of privacy threats: data level and model level, as shown in Fig.~\ref{fig:PrivacyThreatType}. More details are described next.

\subsection{Data-Level Privacy Threats} \label{subsec:datal}

Data-level privacy threats aim at the information contained in published data, such as raw data, identity, statistical properties, etc.

The input part is usually separated from the computation part in practical BCIs. When raw or processed data have been transmitted to the computation part, the input part may no longer be able to control their usage and storage, which may lead to privacy leakage. When the data owner publishes data publicly, there are two possible privacy attack approaches to steal sensitive information of the users, namely, identification attack and inference attack.

\begin{enumerate}

\item \emph{Identification Attack}: This is a kind of de-anonymization attack, which aims to infer the identity of the source from the data obtained. It has been widely used in face recognition \cite{Wilber2016}, location de-anonymization \cite{Gambs2013}, and social network de-anonymization \cite{Ji2016}.

Oh \emph{et al.}~\cite{Oh2016} showed that a few pictures in social media are enough to realize person identification, even some obfuscated images that have been blurred. In fact, many ML approaches can recognize blurred or distorted facial images~\cite{Wilber2016, McPherson}. For example, McPherson \emph{et al.}~\cite{McPherson} empirically demonstrated that artificial neural networks can discover hidden information (e.g., faces, objects, handwritten digits) from images protected by various forms of obfuscation, e.g., mosaicing, blurring, and P3~\cite{Ra2013} (privacy-preserving photo sharing, which encrypts significant JPEG coefficients to make images unrecognizable by humans).

Not only facial images, but also geolocated data can leak identity information. Gambs \emph{et al.}~\cite{Gambs2013} built a mobility Markov chain from the mobility traces in the training set, designed two distance metrics to quantify the similarity between two such chains, and combined them to re-identify users in an anonymized geolocated dataset.

Li \emph{et al.}~\cite{Li2020} proposed an approach to de-anonymize heterogeneous social networks. It first uses the network graph structure to reduce the size of the candidate set, and then performs user identification with the user's profile.

\item \emph{Inference Attack}: This attack performs reasoning or learning on data to obtain hidden sensitive information, which may not be obviously available in the original data.

Feng and Jain~\cite{Feng2011} realized fingerprint reconstruction by converting the compact minutiae representation of a fingerprint into a phase representation, and then a grayscale representation. They showed that the reconstructed grayscale image is similar to the original fingerprint, indicating that it is possible to attack a commercial fingerprint recognition system. Cao and Jian~\cite{Cao2015} further showed that using prior knowledge of the fingerprint ridge structure can achieve better attack performance.

Al-Rubaie and Chang~\cite{AlRubaie2016} used the user's authentication profile or test results of the synthesized sample to reconstruct the raw gesture data, and showed that their approach can be used to attack gesture-based active authentication systems on mobile devices.
\end{enumerate}

To improve the accuracy of BCIs, transfer learning~\cite{Lotte2018,drwuTLBCI2021} has been frequently used, which currently often uses raw data from existing users to facilitate the calibration for a new user. As described in Section~\ref{sub:information}, data in BCIs contain various private information, thus efforts should be made to ensure their security, or new transfer learning approaches that do not need access to other users' raw data~\cite{Liang2020} should be developed.

\subsection{Model-Level Privacy Threats}  \label{subsec:modell}

A proprietary ML model may also be the target for attack, where the attacker aims to extract private information including its structure, parameters, training algorithms, training data, etc.

There are different types of model-level privacy attacks:
\begin{enumerate}
\item \emph{Model Extraction Attack}, which continuously uses the attacker's data to query the model to obtain its outputs, and then uses these input-output pairs to steal model attributes~\cite{Tramer2016} (model structure, hyper-parameters, parameters, training data, etc.) or to imitate model functionalities~\cite{CorreiaSilva2018}. For example, Tram\`{e}r \emph{et al.}~\cite{Tramer2016} demonstrated that online ML-as-a-service, e.g., BigML and Amazon, is subject to model stealing attacks. They proposed simple and efficient equation-solving attacks to extract target ML models, e.g., logistic regression, neural networks, support vector machine, and decision trees, with near-perfect fidelity. Correia-Silva \emph{et al.}~\cite{CorreiaSilva2018} developed a copycat convolutional neural network to copy a proprietary ML model, e.g., Microsoft Azure Emotion API. They first queried the target ML model with random data and recorded its outputs to create a training dataset, which represents the knowledge of the target ML model; then, they trained a copycat convolutional neural network from it, which usually achieved similar performance as the target ML model.

\item \emph{Model Inversion Attack}, where the attacker tries to infer some sensitive information of the training data from the model~\cite{Song2017}. It can be further divided into two categories: membership inference attack and property inference attack. Membership means whether the training data belongs to the training set of a certain model or not. Membership inference attacks assume that the attacker has black-box access to the model, and its purpose is to infer the membership of a model's training data~\cite{Truex2019}. Shokri \emph{et al.}~\cite{Shokri2017} performed membership inference attack, based on the observation that ML models often behave differently on their training data and new data. Property inference attack mainly deduces properties of the target model's training set. For example, Fredrikson \emph{et al.}~\cite{Fredrikson2015} proposed an attack approach that uses confidence values obtained from predictions and successfully applied it to decision trees for lifestyle surveys and neural networks for facial recognition.
\end{enumerate}

To our knowledge, no privacy attacks targeting the BCI models have been reported yet. However, as mentioned in Section~\ref{subsec:datal}, source-free transfer learning~\cite{Liang2020}, which only requires well-trained models from previous users, has been proposed to avoid using other users' data directly (thus to ensure data-level privacy). It requires models in BCIs to be transferred among different users, and model-level privacy attacks could be performed. Additionally, public release of BCI models may facilitate adversarial attacks to them~\cite{Zhang2019a,drwuCIM2021,drwuALBCI2019,drwuNSR2021,drwuTAR2019,drwuUAP2021}, seriously threatening their security.

\subsection{Privacy Threats in BCIs} \label{subsec:threatBCI}

To our knowledge, Martinovic \emph{et al.}~\cite{Martinovic2012} were the first to demonstrate that private and secret information, such as PIN code, bank information, month of birth, known people, area of residence, may be inferred from EEG signals collected by a consumer-grade BCI device. It is based on the detection of the classical P300 signal, a positive peak in the EEG at about 300 ms after the appearance of a target/oddball stimulus. For example, in detecting whether the user owns a particularly credit card, the user was first asked ``\emph{What is the name of your bank?}" and then images of credit cards were flashed on the screen. A P300 may be detected when the user sees a credit card he/she owns. This attack requires special experimental design and active user collaboration, so it is theoretically significant but easy to be detected.

Neupane \emph{et al.}~\cite{Neupane2017} used ERPs to eavesdrop a user's numeric PINs or textual passwords, i.e., to perform side-channel attacks. They assumed that the attacker can develop a malicious application (APP) to access the victim's EEG signals. It requires a training phase, in which the APP captures the neural patterns of keystrokes to build a classifier. If the attacker does not have access to the victim's EEG signals during the training phase, then the classifier may be trained from the attacker's own EEG signals. They demonstrated that attack accuracies were significantly higher than random guessing, for two EEG headsets (a commodity device, Emotiv EPOC Headset, and a higher-end medical-scale device, B-Alert Headset by Advanced Brain Monitoring) and three keyboards (a physical keyboard, a virtual keyboard, and an ATM style numeric keypad). The most distinguishing feature of their approach is that it ``\emph{is highly surreptitious as it only requires passive monitoring of brain signals, not deliberate, and active strategies that may trigger suspicion and be detected by the user.}" So, it is much more difficult to be detected than the previous approach.

To overcome the limitations of Martinovic \emph{et al.}'s approach~\cite{Martinovic2012}, i.e., the attackers use supraliminal (consciously perceived) stimuli and are thus easily detectable, Frank \emph{et al.}~\cite{Frank2017} proposed a subliminal attack that infers private information by visually probing the victim for less than 13.3 milliseconds (to avoid eliciting cognitive perception). The key idea was ``\emph{to show the visual stimuli within the screen content that the user expects to see, but for a duration that is too short for conscious perception (several milliseconds), yet still sufficient to result in activation of certain parts of user's brain detectable by an attacker.}" They demonstrated that, by carefully designing the short visual stimuli, the victim may not be able to notice that he/she is being probed, but the attacker can reduce the entropy of guessing the victim's private information by over 20\%.

Kong \textit{et al.} \cite{Kong2018} demonstrated the possibility of using EEG signals of different BCI paradigms to reveal the user's identity. They assumed that the background EEG in BCIs is user-specific, and proposed a maximum correntropy criterion and a rational quadratic kernel to extract it for person identification. Experiments on four datasets from different BCI paradigms, including Graz dataset A \cite{Brunner2008} on motor imagery (MI), a P300 dataset \cite{Riccio2013} on ERP, SJTU Emotion EEG Dataset (SEED) \cite{Zheng2015} on emotion recognition, and a self-collected multi-task BCI dataset, all achieved over 90\% person identification accuracy.

H\"oeller \emph{et al.}~\cite{Hoeller2018} showed that EEG-biometric templates could threaten user privacy. Experiments on 60 healthy subjects and neurological patients revealed that EEG-biometric templates can be used to classify the subjects by sex, neurological diagnoses, age, atrophy of the brain, and intake of neurological drugs, with accuracy in the range of 70-86\%

Bellman \emph{et al.}~\cite{Bellman2018} tried to classifying unaware facial recognition by consumer-grade BCI devices. During the two-day training period, images of so-called ``ordinary" people were shown to the participants. The experiment included four phases. In Phase 1, participants viewed a total of 162 images, among which 20 pictures were selected and repeated three times each so that the participants would remember them (unaware recognitions). Next day, in Phase 2, 10 of the selected 20 images were randomly selected, combined with 92 new images, and presented to the participants. They assumed that unaware recognition was ``\emph{implicitly learned in the previous day's phase, but the time between the two day's session would have caused participants to forget the faces at an aware level, allowing for an unaware recognition to them in this phase.}" Right away, Phase 3 took place, where participants were asked to remember an image and the image was repeated 20 times while many one-second images were shown. This phase was used to measure aware recognition of the participants. EEG signals were recorded in Phases 2 and 3. In Phase 4, they preprocessed the recorded signals and determined whether an image were subliminally recognized or not. Their experimental results showed high classification accuracies, i.e., it is easy for attackers to determine whether the subject has seen selected faces.

Landau \emph{et al.}~\cite{Landau2020} used a variety of ML approaches, such as convolutional neural networks and $k$-nearest neighbors, to analyze resting-state EEG signals to infer important personality traits and cognitive abilities. Their experiments achieved 73\% classification accuracy when classifying personality traits. They also found that the executive function performance and resting-state EEG signals have significant statistical correlation, indicating the possibility to infer cognitive abilities.

As shown in Table~\ref{tab:ThreatsBCI}, existing privacy attack approaches in BCIs mainly focus on the data-level privacy. The target is often private personal information, such as personal account, personal preferences, and physical state. Source-free transfer learning~\cite{Liang2020} can protect data-level privacy in transfer learning, by requiring only well-trained models from previous users, instead of their original data. However, models trained for BCI tasks contain rich information and have important commercial values. Therefore, directly releasing the models or allowing others to query them freely introduces privacy threats. Additionally, BCI models are the target of adversarial attacks~\cite{Zhang2019a,drwuCIM2021,drwuALBCI2019,drwuNSR2021,drwuTAR2019,drwuUAP2021}, which can manipulate the output of the BCI systems, e.g., change the output of a BCI speller from ``Y" to ``N". Leakage of the BCI models will make adversarial attacks much easier, leading to serious security problems.

\begin{table*}[htpb]  \centering  \renewcommand{\arraystretch}{1.1} \setlength{\tabcolsep}{1.5mm}
    \caption{Existing privacy threats in BCIs.}  \label{tab:ThreatsBCI}
    \begin{tabular}{c|cccc|c|c}  \hline
    \multirow{3}{*}{Reference}&\multicolumn{4}{c|}{Private Information Type}&\multirow{3}{*}{BCI Paradigm}&\multirow{3}{*}{Privacy Attack Performance}\\  \cline{2-5}
    &Personal&Personal &Physical&Commercial&&\\
    &Account&Preference&State&Model&&\\ \hline
   \cite{Martinovic2012}&\checkmark&&&&ERP&Significantly better than random attack \\
   \cite{Frank2017}&\checkmark&&&&ERP&20.84\% guess entropy reduction in subliminal probing\\
   \cite{Neupane2017}&\checkmark&&&&ERP&20.5-47.5\% true positive rate in four tasks\\
   \cite{Hoeller2018}&&\checkmark&&&-&60-86\% accuracy in nine classification tasks\\
   \cite{Kong2018}&\checkmark&&&&MI, ERP, emotion recognition&Over 90\% person identification accuracy\\
   \cite{Bellman2018}&\checkmark&&&&ERP&Up to 0.773 F-score \\
   \cite{Landau2020}&&\checkmark&\checkmark&&Resting-state EEG&73\% personality trait classification accuracy\\ \hline
    \end{tabular}
  \end{table*}

\section{Possible Solutions to Privacy Threats in BCIs} \label{sub:protection}

The previous section introduced potential privacy threats in BCIs. In this section, we describe possible privacy protection approaches to defend against these threats.

Anonymization and data sanitization can reduce the privacy loss caused by attacks. Yu \textit{et al.}~\cite{Yu2020} proposed a blockchain based medical research support platform against COVID-19, which uses a pseudonym mechanism to anonymize patients' data for privacy protection. Iwendi \textit{et al.}~\cite{Iwendi2020} proposed an N-sanitized de-identification framework to protect the privacy of unstructured medical datasets, which includes detection and sanitization of personal health identifiers (e.g., name, email address), detection and removal of negative assertions [e.g., human immunodeficiency virus (HIV) negative], detection and sanitization of sensitive terms (e.g., symptoms, tests, disease, treatment), and random sampling of the document to manipulate the adversary belief and achieve semantic privacy protection.

\begin{figure}[htpb]   \centering
  \includegraphics[width=\linewidth,clip]{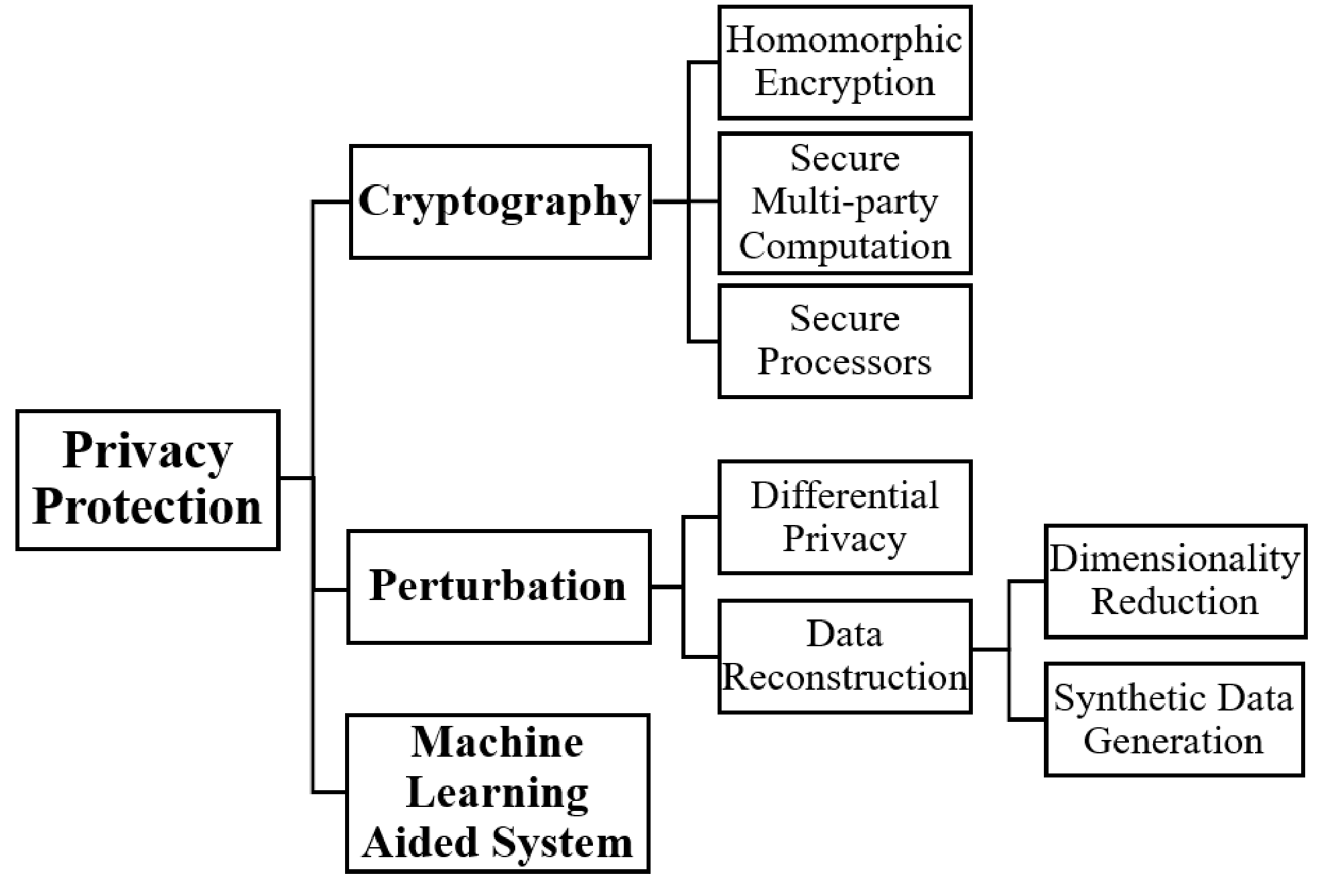}
  \caption{Privacy protection approaches.}   \label{fig:PrivacyProtection}
  \end{figure}

Next, we introduce additional approaches for reducing the possibility of inferring private information from EEGs, which can be divided into three categories: cryptography, perturbation, and ML aided systems, as shown in Fig.~\ref{fig:PrivacyProtection}.

\subsection{Cryptography}  \label{subsec:cryptography}

Cryptography is an important strategy against privacy threats in privacy-preserving ML. However, common ML approaches may not be able to deal with encrypted data. To overcome this limitation, Bost \emph{et al.}~\cite{Bost2015} designed efficient protocols for many core operations on encrypted data in many classification algorithms, such as comparison, argmax and dot product. Dowlin \emph{et al.}~\cite{Dowlin2016} proposed CrytoNets to train neural networks on encrypted data to make encrypted predictions.

There are three categories of cryptography-based approaches: homomorphic encryption (HE), secure multi-party computation (SMC), and secure processors. Since the data have been encrypted via HE protocols, they can be securely computed by outsourced parties via SMC. Both operations can be implemented on secure processors for further protection.

\emph{Homomorphic Encryption}: HE is an encryption strategy that makes operations on encrypted data, such as addition and multiplication, feasible. Additive HE only enables addition on encrypted data, and can be used to construct extreme learning machines~\cite{Kuri2017}. Nonlinear calculations can be approximated by additive HE, but the learning performance may be affected. Full HE was proposed to enable arbitrary operations on encrypted data, which was used in~\cite{Sun2020} to construct hyperplane decision-based classifiers, Na\"{i}ve Bayesian classifiers, and decision trees. However, it is time-consuming to perform full HE.

\emph{Secure Multi-party Computation}: SMC is a cryptographic protocol that distributes the computation across multiple parties, and no individual party can see other parties' data. It is often used when multiple parties wish to train a model together, where each party holds a share of the training data. Common strategies used in SMC are garbled circuits~\cite{Yao1986} and secret sharing~\cite{Ahlswede1993}. Nikolaenko \emph{et al.}~\cite{Nikolaenko2013} achieved privacy-preserving ridge regression via Yao garbled circuits~\cite{Yao1986}, which was also combined with HE for better security. Secret sharing seeks to distribute a secret among many parties, one share for one party~\cite{Ahlswede1993}. A single share does not have any information; only by combining them all-together can one get the secret. 

\emph{Secure processors}: Secure processors perform computations on trusted hardware, e.g., Intel Software Guard Extensions (SGX). Ohrimenko \emph{et al.}~\cite{Ohrimenko2016} implemented privacy-preserving multi-party machine learning on trusted SGX processors, assuming that ``\emph{the adversary is unable to physically open and manipulate the SGX processor chips that run the machine learning computation}". Secure processors usually do not support high-level languages, which is inconvenient for data scientists without security knowledge to use. Shaon \emph{et al.}~\cite{Shaon2017} provided Python/Matlab-like high-level language supports to help data scientists complete data analytic tasks on secure processors.

It is common to use other users' data to help calibrate a BCI system for a new user~\cite{drwuTLBCI2021}. Cryptography can be used to protect other users' data privacy, e.g., SMC-based linear regression has been proposed~\cite{drwuBigData2018,Agarwal2019}. Since cryptography usually has high computational complexity, and BCI data usually have high-dimensionality, the training efficiency may be seriously impacted. How to improve the efficiency of cryptography-based algorithms deserves future research.

\subsection{Perturbation} \label{subsec:perturbation}

Perturbation can be used to resist attacks, by adding certain noise or transforming the original data, while maintaining the data utility for downstream tasks. For example, Zhang \emph{et al.}~\cite{Zhang2018} obfuscated the original samples by adding random noise, or augmented samples, to the training set. Therefore, sensitive information about the properties of individual samples, or statistical properties of the training set, is hidden, but the dataset can still maintain high utility. Main strategies in perturbation can be partitioned into two categories: differential privacy and data reconstruction.

\emph{Differential Privacy}: Differential privacy describes the patterns of groups within a dataset while withholding information about individuals in the dataset~\cite{Zhu2017}. Common strategies to implement differential privacy are Laplace mechanism~\cite{Dwork2011} and exponential mechanism~\cite{McSherry2007}.

Many ML algorithms have been combined with differential privacy technology for privacy protection. For example, Rubinstein \emph{et al.}~\cite{Rubinstein2009} proposed private support vector machine based on differential privacy for finite-dimensional feature mapping and infinite-dimensional feature mapping with translation-invariant kernels. They studied the trade-off between utility and the guaranteed differential privacy level. Unfortunately, their results showed that there is no mechanism to maximize both of them simultaneously. Chaudhuri \emph{et al.}~\cite{Chaudhuri2011} proposed to approximate empirical risk minimization under differential privacy. They compared objective perturbation, which adds noise to the objective function of the empirical risk before optimization, and output perturbation, which adds noise to the output of the empirical risk. Their experimental results showed that the former is better.

However, traditional differential privacy requires a trusted third-party data center to store and publish private data, which is unpracticable in many real-world applications. Therefore, local differential privacy emerges~\cite{Xiong2020}, which allows users to individually protect personal sensitive information. However, since each user applies his/her own randomized mechanism to perturb the private data before publishing them to an un-trusted server, the data utility may degrade. Distributed Differential Privacy makes a trade-off between the traditional differential privacy and  local differential privacy to guarantee both differential privacy and utility, by requiring each group (multiple users) to use a random algorithm instead of each user in local differential privacy~\cite{Shi2011}.

Adversarial perturbation is another popular strategy to provide privacy guarantees in differential privacy. Liu \emph{et al.}~\cite{Liu2017} proposed to automatically detect objects in an image and craft specific adversarial examples to protect the privacy of social media images without affecting the visual quality of human eyes. More specifically, they used adversarial perturbation to fool the region proposal process in objection detection, so that no proposal boxes can be classified in the next procedure. They also found that the crafted adversarial examples for one deep neural network can be useful to others.

\emph{Data Reconstruction}: Privacy and utility are two important issues for data publishing. Ideally, information related to downstream tasks should be maintained, and irrelevant sensitive information should be eliminated. For example, EEG signals used in a clinical diagnostic system can be processed to retain disease-related information but not sensitive information such as identity~\cite{Pascual2020}. Data reconstruction can achieve this purpose, via two popular strategies: dimensionality reduction and synthetic data generation.

Dimensionality reduction projects the original data onto a lower dimensional space, which is impossible for the adversary to reconstruct the original data. In this way, dimensionality reduction may eliminate the sensitive information while maintaining the utility for downstream tasks. Liu \emph{et al.}~\cite{Liu2006} found that random orthogonal transformation cannot resist attacks performed by Independent Component Analysis. Therefore, they proposed to project the multiplicative data onto a random subspace, where the original data are hidden and distance-based statistical properties are preserved. Alotaibi \emph{et al.}~\cite{Alotaibi2012} used a nonlinear dimensionality reduction approach, non-metric multidimensional scaling, to perturb the data for privacy preserving while maintaining the utility for data clustering. However, if the dimensionality reduction mapping can be learned from the data, then it still needs to be differentially private~\cite{Niinimaeki2019}.

Synthetic data generation is another popular strategy. Generative models are important for data owners to publish their datasets when they want to protect data privacy and maintain data utility simultaneously. GANs have been used for generating natural images~\cite{Denton2015}, simulating mobility trajectories~\cite{Kulkarni2018}, and so on. Recently, there have been several approaches using GANs to publish private data~\cite{SamsonCheung2018,Acs2019}. For example, Cheung \emph{et al.}~\cite{SamsonCheung2018} proposed to use generated facial images to train face classifiers. First, a GAN model was trained in the trusted domain. Then, its parameters, or the generated samples through the model, were sent to a public domain for downstream tasks. Their experimental results showed little classification performance degradation, compared with standard classification approaches.

Synthesized EEG data have been used in BCIs~\cite{Debie2020,Pascual2020}. Downstream model training would be affected if the fidelity of these synthesized data is low. Similarly, perturbation methods based on differential privacy or dimensionality reduction also lead to the loss of task-related information. A possible solution is to disentangle the task-related component and the task-irrelevant component of the data, so that the perturbation is only applied to the task-irrelevant component to minimize the leakage of private information while maintaining the data utility.

\subsection{Machine Learning Aided Systems} \label{subsec:MLaided}

ML aided systems aim to help people better understand privacy policies, and inform them about the privacy risks when making privacy decisions. They have been used to assess the privacy risk~\cite{Sion2019}, alert people when publishing sensitive information~\cite{Orekondy2018}, facilitate the understanding of privacy policies~\cite{Liu2020}, and recommend privacy settings according to users' preferences~\cite{Wijesekera2017}.

For example, Fu \emph{et al.}~\cite{Fu2019} tried to answer the question ``\emph{is the resource access indicated by APP foreground?}" and make the APP foreground interaction and background access consistent. They checked the contextual integrity in the APP foreground interaction to detect malicious APPs. In the offline phase, they trained a model to learn the relationship between sensitive API calls and the corresponding foreground windows. In the online phase, they used the model to detect malicious API calls and personalized the model with the user's feedback.

Many users do not know how their data are transmitted and used in BCIs. ML aided systems can be developed to monitor the data transmission and usage, and alert the user when there are privacy threats. This may be necessary to meet legal requirements, and ensure the trust of users.

\subsection{Existing Solutions in BCIs} \label{subsec:solutionBCI}

There have been some cryptography-based approaches for privacy protection in BCIs.

Robinson and Varghese~\cite{Robinson2016} encrypted raw EEG data and provided only task-related features to researchers for privacy protection. They focused on features related to three kinds of neurological disorders, i.e., epilepsy, Alzheimer's disease, and Schizophrenia. Different applications are corresponding to different feature extraction algorithms. Specifically, they used discrete wavelet transform to extract features for epilepsy and Alzheimer's disease, and an autoregressive method for Schizophrenia. In the data transfer phase, each user encrypted his/her raw EEG data using the Advanced Encryption Standard algorithm and sent the encrypted data to other participants. A researcher can request access to the task-related features, which will be decrypted under the control of the user.

Agarwal \emph{et al.}~\cite{drwuBigData2018} were the first to apply commodity-based SMC to EEG data for driver drowsiness estimation. They performed linear regression over EEG signals from 15 users in a fully privacy-preserving manner, so that each individual's EEG signals are not revealed to anyone else. The results were identical to those of the un-encrypted case, with reasonable computational cost.

Agarwal \emph{et al.}~\cite{Agarwal2019} further proposed an SMC-based privacy-preserving ML approach. They considered the scenarios with and without calibration data in the target party separately. They developed an SMC-based linear regression algorithm for the scenario that there are no calibration data in the target party. In the training process, a trusted third-party distributes random numbers to each source party for following secure computations. Each source party individually computes the covariance matrix and other required variables on its own data share, and then jointly computes other variables based on secure matrix computation protocols and random numbers produced by the trusted third-party. Finally, each source party obtains a share of the linear regression model. In the inference phase, the target party first produces shares of its input data and distributes them to the source parties respectively. Then, all source parties jointly compute the result and obtain their own share of the result. Finally, all source parties send their shares to the target party, which then aggregates the received shares to obtain the final result.

Generated EEG data can be shared without privacy leakage to some extent. Debie \emph{et al.}~\cite{Debie2020} tried to use GANs to generate synthetic EEG data for downstream tasks rather than directly sharing the real data. They claimed that the generated synthetic EEG data can be similar to the real data but do not contain personal sensitive information. There are two components in the GAN model: a generator $G$ and a discriminator $D$. The $G$ network was used to generate synthetic EEG data and the $D$ network tried to distinguish the generated data from real ones. In addition, the $D$ network was trained via a differentially private stochastic gradient descent approach for further privacy protection. Their experimental results showed that the generated EEG data perform comparably with the real data.

Pascual \emph{et al.}~\cite{Pascual2020} proposed EpilepsyGAN to synthesize ictal seizure EEG signals. The real ictal EEG signals are hard to record because the onset of seizure is rare and unpredictable. Real ictal EEG signals may be used for patient re-identification, threatening patients' privacy. The synthetic ictal EEG signals can deal with these two issues. They used inter-ictal EEG signals to generate ictal EEG signals for a new patient, and then the synthetic ictal EEG signals for downstream tasks for privacy protection. Their experimental results showed that the data utility of the synthetic signals may be even higher than the real signals, and patient re-identification from the synthetic signals was much more challenging.

Popescu \textit{et al.} \cite{Popescu2021} proposed ``\emph{an encoding method that enables typical HE schemes to operate on real-valued numbers of arbitrary precision and size}." The training time increased, but the inference can be performed very fast. Evaluations on EEG-based seizure detection and prediction of predisposition to alcoholism showed ``\emph{the prediction performance of the models operating on encoded and encrypted data is comparable to that of standard models operating on plaintext data.}"

As summarized in Table~\ref{tab:Solutions}, most existing privacy-preserving approaches in BCIs focused on data-level privacy, based on cryptography and perturbation. Particularly, \cite{drwuBigData2018,Agarwal2019,Debie2020} use cryptography or perturbation to process the data and transmit the processed data among users, which ensures that other users cannot infer private information from these data. At the same time, because they eliminate the need to transfer models among users, they can also resist model-level privacy attacks to a certain degree.

Additionally, since EEG can be viewed as a type of medical image, approaches proposed for medical image privacy protection may also be used for BCIs. For example, Montenegro \textit{et al.} \cite{Montenegro2021} proposed to use a Conditional Variational Autoencoder to reconstruct the original images with identity replacement, which is similar to using GAN in EpilepsyGAN to reconstruct seizure EEG signals.

To our knowledge, ML aided systems have not been studied in privacy-preserving BCIs yet. There are plenty of research opportunities.

\begin{table}[htpb]  \centering  \renewcommand{\arraystretch}{1.1} \setlength{\tabcolsep}{1.5mm}
  \caption{Existing privacy protection solutions in BCIs.}  \label{tab:Solutions}
  \begin{tabular}{c|cc|ccc}  \hline
  \multirow{3}{*}{Reference}&\multicolumn{2}{c|}{Target}&\multicolumn{3}{c}{Type}\\ \cline{2-6}
  &Data&Model&\multirow{2}{*}{Cryptography}&\multirow{2}{*}{Perturbation}&ML Aided\\
  &Level&Level&&&System\\  \hline
  ~\cite{Robinson2016}&\checkmark&&\checkmark&\checkmark&\\
  ~\cite{drwuBigData2018} &\checkmark&\checkmark&\checkmark&&\\
  ~\cite{Agarwal2019}&\checkmark&\checkmark&\checkmark&&\\
   ~\cite{Pascual2020}&\checkmark&&&\checkmark&\\
 ~\cite{Debie2020}&\checkmark&\checkmark&&\checkmark&\\
    ~\cite{Popescu2021}&\checkmark&&\checkmark&&\\ \hline
  \end{tabular}
\end{table}

\section{Challenges of Privacy-Preserving ML in BCIS} \label{sub:challenge}

We have introduced the privacy threats and the development of privacy-preserving ML approaches in BCIs in previous sections. There are many challenges on privacy-preserving ML in BCIs, including cross-subject variations in EEG signals, balance of utility and privacy, balance of computation cost and privacy, and evaluation. They are discussed in detail next.

\subsection{Cross-Subject Variations in EEG Signals}  \label{subsec:transfer}

EEG is the most widely used input signal in BCIs~\cite{Schomer2017}. However, it suffers from a low signal-to-noise ratio, poor spatial resolution, and high cross-subject variations~\cite{Lotte2018}. Many signal processing techniques have been proposed to improve the quality of EEG signals, whereas it is much more challenging to accommodate cross-subject variations. One promising solution to reducing the variations is transfer learning~\cite{Pan2010,drwuTLBCI2021,drwuEA2020}, which utilizes auxiliary data or knowledge from previous subjects/tasks (\emph{source domains}) to facilitate the calibration for a new subject/task (\emph{target domain}). Since many transfer learning approaches require access to source domains' data, which may contain privacy information of the source subjects, there are privacy concerns.

Source-free transfer learning~\cite{Liang2020} seems to be a promising solution to address this issue. It only requires access to the source domain models rather than the original data. Although there are privacy attacks targeting published models, less private information can be stolen from models than data. Perturbation-based approaches may be difficult to use here, due to the performance degradation, although they can provide complete privacy guarantees. Furthermore, Zhang \emph{et al.}~\cite{Zhang2021} proposed to realize knowledge transfer with a black-box source model.

Federated learning~\cite{Rahman2020} is another promising solution. Since it aims to train a model via multi-source data resided in distributed devices, it can be combined with transfer learning to cope with cross-subject variations. Liu \emph{et al.}~\cite{Liu2020a} proposed a secure federated transfer learning framework. Ju \emph{et al.}~\cite{Ju2020} further combined this framework with the covariance matrices of EEG signals for privacy-preserving transfer learning.

Despite that parameter transfer and federated transfer learning are promising solutions, they are not resistant to all privacy attacks, especially model-targeted attacks. Therefore, future research may consider combining encryption or perturbation approaches with them for improved robustness.

\subsection{Balance of Utility and Privacy} \label{subsec:utility}

Some privacy protection approaches degrade the data utility or model performance. For example, differential privacy realizes privacy protection by adding certain disturbances to data, models, or intermediate variables, in the training process. If the level of privacy protection is too demanding, then model training may not converge, or the converged model may not perform satisfactorily. Therefore, when developing privacy protection approaches in BCIs, their impact on the utility should also be carefully evaluated.


\subsection{Balance of Computation Cost and Privacy} \label{subsec:cost}

NeuroCrypt \cite{Senanayake2021}, which performs ML over encrypted distributed neuro-imaging data, is emerging. However, cryptography-based approaches are computationally demanding. For instance, HE can achieve homomorphism between plain-text operations and cipher-text operations; however, it needs to map the plain-text space onto a cipher-text space, which is usually high-dimensional. The calculation in such a space is very time-consuming, and the encryption process is very computationally intensive. In addition, in order to achieve a higher level of privacy protection, it is necessary to continuously increase the length of the cipher-text, which further increases the computational cost.

Time-consuming privacy protection approaches may not be suitable for real-time online BCI systems. Therefore, there is a need to balance computational cost and privacy protection in BCIs.

%

\subsection{Evaluation}  \label{subsec:evaluation}

As there are many different privacy-preserving approaches, it is necessary to benchmark their performances. Unfortunately, this is not easy. For example, we cannot directly compare the privacy guarantee provided by HE with a certain level of differential privacy. Nevertheless, being able to compare the privacy-preserving capability of different approaches in BCIs is desirable.

\section{Conclusions and Future Directions} \label{sub:conclusion}

BCIs have been widely used in medical diagnosis, rehabilitation, education, entertainment, and so on. Most research so far focused on making BCIs more accurate and reliable, but much less attention has been paid to their privacy. Input signals in BCIs, e.g., EEG, contain rich privacy information, and the developed machine learning model is usually proprietary. Data and model transmission among different parties during BCI development and applications may incur significant privacy threats, and hence privacy protection in BCIs must be considered. This paper presents a contemporary and comprehensive review on privacy-preserving BCIs. It describes potential privacy threats and protection strategies in BCIs, and points out several challenges in developing privacy-preserving BCIs.

Future research directions include:
\begin{enumerate}
\item Machine learning approaches that can handle cross-subject variations and data-level privacy threats simultaneously. Source-free transfer learning and federated learning are two promising solutions, but so far their applications in BCIs are very limited.
\item Approaches to disentangle the task-related component and the task-irrelevant component of the BCI data, so that perturbations can be applied to only the task-irrelevant component to reduce the risk of privacy leakage while maintaining the data utility.
\item Reducing the computation and communication cost of (encryption-based) privacy protection algorithms, so that they can scale to high dimensional data and large number of users in BCIs.
\item An index to quantify the degree that user privacy is protected, so that different algorithms can be benchmarked.
\end{enumerate}

We hope this review can attract more research interests to privacy-preserving BCIs, and hence facilitates a smoother transition of BCIs to the market.




\begin{thebibliography}{100}
\providecommand{\url}[1]{#1}
\csname url@samestyle\endcsname
\providecommand{\newblock}{\relax}
\providecommand{\bibinfo}[2]{#2}
\providecommand{\BIBentrySTDinterwordspacing}{\spaceskip=0pt\relax}
\providecommand{\BIBentryALTinterwordstretchfactor}{4}
\providecommand{\BIBentryALTinterwordspacing}{\spaceskip=\fontdimen2\font plus
\BIBentryALTinterwordstretchfactor\fontdimen3\font minus
  \fontdimen4\font\relax}
\providecommand{\BIBforeignlanguage}[2]{{%
\expandafter\ifx\csname l@#1\endcsname\relax
\typeout{** WARNING: IEEEtran.bst: No hyphenation pattern has been}%
\typeout{** loaded for the language `#1'. Using the pattern for}%
\typeout{** the default language instead.}%
\else
\language=\csname l@#1\endcsname
\fi
#2}}
\providecommand{\BIBdecl}{\relax}
\BIBdecl

\bibitem{Gu2020}
X.~Gu, Z.~Cao, A.~Jolfaei, P.~Xu, D.~Wu, T.-P. Jung, and C.-T. Lin,
  ``{EEG}-based brain-computer interfaces ({BCIs}): A survey of recent studies
  on signal sensing technologies and computational intelligence approaches and
  their applications,'' \emph{IEEE/ACM Trans. on Computational Biology and
  Bioinformatics}, vol.~18, no.~5, pp. 1645--1666, 2021.

\bibitem{Li2020a}
Y.~{Li}, Y.~{Liu}, W.-G. Cui, Y.-Z. Guo, H.~{Huang}, and Z.-Y. Hu, ``Epileptic
  seizure detection in {EEG} signals using a unified temporal-spectral
  squeeze-and-excitation network,'' \emph{IEEE Trans. on Neural Systems and
  Rehabilitation Engineering}, vol.~28, no.~4, pp. 782--794, 2020.

\bibitem{Zeng2008}
F.~{Zeng}, S.~{Rebscher}, W.~{Harrison}, X.~{Sun}, and H.~{Feng}, ``Cochlear
  implants: System design, integration, and evaluation,'' \emph{IEEE Reviews in
  Biomedical Engineering}, vol.~1, pp. 115--142, 2008.

\bibitem{Eugster2016}
M.~J. Eugster, T.~Ruotsalo, M.~M. Spap\'{e}, O.~Barral, N.~Ravaja, G.~Jacucci,
  and S.~Kaski, ``Natural brain-information interfaces: Recommending
  information by relevance inferred from human brain signals,''
  \emph{Scientific Reports}, vol.~6, p. 38580, 2016.

\bibitem{Ienca2017}
M.~Ienca, J.~Fabrice, B.~Elger, M.~Caon, A.~S. Pappagallo, R.~W. Kressing, and
  T.~Wangmo, ``Intelligent assistive technology for {Alzheimers} disease and
  other dementias: a systematic review,'' \emph{Journal of Alzheimer's
  Disease}, vol.~56, no.~4, pp. 1301--1340, 2017.

\bibitem{Neupane2019}
A.~{Neupane}, K.~{Satvat}, M.~{Hosseini}, and N.~{Saxena}, ``Brain hemorrhage:
  When brainwaves leak sensitive medical conditions and personal information,''
  in \emph{Proc. 17th Int'l Conf. on Privacy, Security and Trust}, Fredericton,
  Canada, Aug. 2019, pp. 1--10.

\bibitem{Palazzo2017}
S.~{Palazzo}, C.~{Spampinato}, I.~{Kavasidis}, D.~{Giordano}, and M.~{Shah},
  ``Generative adversarial networks conditioned by brain signals,'' in
  \emph{Proc. IEEE Int'l' Conf. on Computer Vision}, Venice, Italy, Oct. 2017,
  pp. 3430--3438.

\bibitem{Frank2017}
M.~Frank, T.~Hwu, S.~Jain, R.~T. Knight, I.~Martinovic, P.~Mittal, D.~Perito,
  I.~Sluganovic, and D.~Song, ``Using {EEG}-based {BCI} devices to subliminally
  probe for private information,'' in \emph{Proc. Workshop on Privacy in the
  Electronic Society}, Dallas, TX, Oct. 2017, pp. 133--136.

\bibitem{Bellman2018}
C.~{Bellman}, M.~{Vargas Martin}, and S.~{MacDonald}, ``On the potential of
  data extraction by detecting unaware facial recognition with brain-computer
  interfaces,'' in \emph{Proc. IEEE Int'l Conf. on Cognitive Computing}, San
  Francisco, CA, Jul. 2018, pp. 99--105.

\bibitem{Landau2020}
O.~Landau, A.~Cohen, S.~Gordon, and N.~Nissim, ``Mind your privacy: Privacy
  leakage through {BCI} applications using machine learning methods,''
  \emph{Knowledge-Based Systems}, vol. 198, p. 105932, 2020.

\bibitem{QianQianLi2015}
Q.~Li, D.~Ding, and M.~{Conti}, ``Brain-computer interface applications:
  Security and privacy challenges,'' in \emph{Proc. IEEE Conf. on
  Communications and Network Security}, Florence, Italy, Sep. 2015, pp.
  663--666.

\bibitem{Takabi2016}
H.~{Takabi}, A.~{Bhalotiya}, and M.~{Alohaly}, ``Brain computer interface
  ({BCI}) applications: Privacy threats and countermeasures,'' in \emph{Proc.
  IEEE 2nd Int'l Conf. on Collaboration and Internet Computing}, Pittsburgh,
  PA, Nov. 2016, pp. 102--111.

\bibitem{Yuste2017}
R.~Yuste, S.~Goering, G.~Bi, J.~M. Carmena, A.~Carter, J.~J. Fins, P.~Friesen,
  J.~Gallant, J.~E. Huggins, J.~Illes \emph{et~al.}, ``Four ethical piorities
  for neurotechnologies and {AI},'' \emph{Nature}, vol. 551, no. 7679, pp.
  159--163, 2017.

\bibitem{Ienca2018}
M.~Ienca, P.~Haselager, and E.~J. Emanuel, ``Brain leaks and consumer
  neurotechnology,'' \emph{Nature Biotechnology}, vol.~36, no.~9, pp. 805--810,
  2018.

\bibitem{DelPozoBanos2015}
M.~DelPozo-Banos, C.~M. Travieso, C.~T. Weidemann, and J.~B. Alonso, ``{EEG}
  biometric identification: A thorough exploration of the time-frequency
  domain,'' \emph{Journal of Neural Engineering}, vol.~12, no.~5, p. 056019,
  2015.

\bibitem{RuizBlondet2014}
M.~Ruiz-Blondet, N.~Khlaifian, B.~Armstrong, Z.~Jin, K.~Kurtz, and S.~Laszlo,
  ``Brainprint: Identifying unique features of neural activity with machine
  learning,'' in \emph{Proc. Annual Meeting of the Cognitive Science Society},
  Quebec City, Canada, Jul. 2014.

\bibitem{Martinovic2012}
I.~Martinovic, D.~Davies, M.~Frank, D.~Perito, T.~Ros, and D.~Song, ``On the
  feasibility of side-channel attacks with brain-computer interfaces,'' in
  \emph{Proc. 21st {USENIX} Security Symposium}, Bellevue, WA, Aug. 2012, pp.
  143--158.

\bibitem{Neupane2017}
A.~Neupane, M.~L. Rahman, and N.~Saxena, ``{PEEP}: Passively eavesdropping
  private input via brainwave signals,'' in \emph{Proc. Financial Cryptography
  and Data Security}, Sliema, Malta, Apr. 2017, pp. 227--246.

\bibitem{Choi2018}
G.~{Choi}, S.~{Choi}, and H.~{Hwang}, ``Individual identification based on
  resting-state {EEG},'' in \emph{Proc. 6th Int'l Conf. on Brain-Computer
  Interface}, Gangwon, Korea, Jan. 2018, pp. 1--4.

\bibitem{Knutson2007}
K.~M. Knutson, L.~Mah, C.~F. Manly, and J.~Grafman, ``Neural correlates of
  automatic beliefs about gender and race,'' \emph{Human Brain Mapping},
  vol.~28, no.~10, pp. 915--930, 2007.

\bibitem{Vecchiato2010}
G.~{Vecchiato}, J.~{Toppi}, F.~{Cincotti}, L.~{Astolfi}, F.~{De Vico Fallani},
  F.~{Aloise}, D.~{Mattia}, S.~{Bocale}, F.~{Vernucci}, and F.~{Babiloni},
  ``Neuropolitics: {EEG} spectral maps related to a political vote based on the
  first impression of the candidate's face,'' in \emph{Proc. Annual Int'l Conf.
  of the IEEE Engineering in Medicine and Biology}, Buenos Aires, Argentina,
  Aug. 2010, pp. 2902--2905.

\bibitem{Jost2014}
J.~T. Jost, H.~H. Nam, D.~M. Amodio, and J.~J.~V. Bavel, ``Political
  neuroscience: The beginning of a beautiful friendship,'' \emph{Political
  Psychology}, vol.~35, no.~S1, pp. 3--42, 2014.

\bibitem{Schreiber2017}
D.~Schreiber, ``Neuropolitics: Twenty years later,'' \emph{Politics and the
  Life Sciences}, vol.~36, no.~2, pp. 114--131, 2017.

\bibitem{Galli2021}
G.~Galli, D.~Angelucci, S.~Bode, C.~D. Giorgi, L.~D. Sio, A.~Paparo, G.~D.
  Lorenzo, and V.~Betti, ``Early {EEG} responses to pre-electoral survey items
  reflect political attitudes and predict voting behavior,'' \emph{Scientific
  Reports}, vol.~11, p. 18692, 2021.

\bibitem{Filippi2010}
M.~Filippi, G.~Riccitelli, A.~Falini, F.~Di~Salle, P.~Vuilleumier, G.~Comi, and
  M.~A. Rocca, ``The brain functional networks associated to human and animal
  suffering differ among omnivores, vegetarians and vegans,'' \emph{PLoS ONE},
  vol.~5, no.~5, pp. 1--9, 2010.

\bibitem{Spape2021}
M.~{Spap\'e}, K.~{Davis}, L.~{Kangassalo}, N.~{Ravaja},
  Z.~{Sovij\"arvi-Spap\'e}, and T.~{Ruotsalo}, ``Brain-computer interface for
  generating personally attractive images,'' \emph{IEEE Trans. on Affective
  Computing}, 2021, in press.

\bibitem{Davis2021}
K.~M. {Davis \uppercase\expandafter{\romannumeral3}}, M.~M.~A. Spap{\'{e}}, and
  T.~Ruotsalo, ``Collaborative filtering with preferences inferred from brain
  signals,'' in \emph{Proc. of the Web Conf.}, Virtual Event, Apr. 2021, pp.
  602--611.

\bibitem{Groves2017}
K.~Groves, S.~Kennett, and H.~Gillmeister, ``Evidence for erp biomarkers of
  eating disorder symptoms in women,'' \emph{Biological Psychology}, vol. 123,
  pp. 205--219, 2017.

\bibitem{Lutz2021}
M.~C. Lutz, R.~Kok, and I.~H.~A. Franken, ``Event-related potential ({ERP})
  measures of error processing as biomarkers of externalizing disorders: A
  narrative review,'' \emph{International Journal of Psychophysiology}, vol.
  166, pp. 151--159, 2021.

\bibitem{Nestor2018}
L.~J. Nestor, E.~McCabe, J.~Jones, L.~Clancy, and H.~Garavan, ``Smokers and
  ex-somkers have shared differences in the neural substrates for potential
  monetary gains and losses,'' \emph{Addiction Biology}, vol.~23, no.~1, pp.
  369--378, 2018.

\bibitem{Zhang2019a}
X.~{Zhang} and D.~{Wu}, ``On the vulnerability of {CNN} classifiers in
  {EEG}-based {BCIs},'' \emph{IEEE Trans. on Neural Systems and Rehabilitation
  Engineering}, vol.~27, no.~5, pp. 814--825, 2019.

\bibitem{Tramer2016}
F.~Tram\`{e}r, F.~Zhang, A.~Juels, M.~K. Reiter, and T.~Ristenpart, ``Stealing
  machine learning models via prediction {APIs},'' in \emph{Proc. 25th USENIX
  Security Symposium}, Austin, TX, Aug. 2016, pp. 601--618.

\bibitem{CorreiaSilva2018}
J.~R. {Correia-Silva}, R.~F. {Berriel}, C.~{Badue}, A.~F. {de Souza}, and
  T.~{Oliveira-Santos}, ``Copycat {CNN}: Stealing knowledge by persuading
  confession with random non-labeled data,'' in \emph{Proc. Int'l Joint Conf.
  on Neural Networks}, Rio, Brazil, Jul. 2018, pp. 1--8.

\bibitem{Papernot2017}
N.~Papernot, P.~McDaniel, I.~Goodfellow, S.~Jha, Z.~B. Celik, and A.~Swami,
  ``Practical black-box attacks against machine learning,'' in \emph{Proc. ACM
  on Asia Conf. on Computer and Communications Security}, Abu Dhabi, United
  Arab Emirates, Apr. 2017, pp. 506--519.

\bibitem{AlRubaie2019}
M.~{Al-Rubaie} and J.~M. {Chang}, ``Privacy-preserving machine learning:
  Threats and solutions,'' \emph{IEEE Security \& Privacy}, vol.~17, no.~2, pp.
  49--58, 2019.

\bibitem{Wilber2016}
M.~J. {Wilber}, V.~{Shmatikov}, and S.~{Belongie}, ``Can we still avoid
  automatic face detection?'' in \emph{Proc. IEEE Winter Conf. on Applications
  of Computer Vision}, Lake Placid, NY, Mar. 2016, pp. 1--9.

\bibitem{Gambs2013}
S.~{Gambs}, M.~{Killijian}, and M.~N. d.~P.~{Cortez}, ``De-anonymization attack
  on geolocated data,'' in \emph{Proc. 12th IEEE Int'l Conf. on Trust, Security
  and Privacy in Computing and Communications}, Melbourne, Australia, Jul.
  2013, pp. 789--797.

\bibitem{Ji2016}
S.~{Ji}, W.~{Li}, M.~{Srivatsa}, and R.~{Beyah}, ``Structural data
  de-anonymization: Theory and practice,'' \emph{IEEE/ACM Trans. on
  Networking}, vol.~24, no.~6, pp. 3523--3536, 2016.

\bibitem{Oh2016}
S.~J. Oh, R.~Benenson, M.~Fritz, and B.~Schiele, ``Faceless person recognition:
  Privacy implications in social media,'' in \emph{Proc. European Conf. on
  Computer Vision}, Amsterdam, The Netherlands, Oct. 2016, pp. 19--35.

\bibitem{McPherson}
R.~McPherson, R.~Shokri, and V.~Shmatikov, ``Defeating image obfuscation with
  deep learning,'' \emph{arXiv preprint arXiv:1609.00408}, 2016.

\bibitem{Ra2013}
M.-R. Ra, R.~Govindan, and A.~Ortega, ``P3: Toward privacy-preserving photo
  sharing,'' in \emph{Proc. 10th USENIX Symposium on Networked Systems Design
  and Implementation}, Lombard, IL, Apr. 2013, pp. 515--528.

\bibitem{Li2020}
H.~{Li}, Q.~{Chen}, H.~{Zhu}, D.~{Ma}, H.~{Wen}, and X.~S. {Shen}, ``Privacy
  leakage via de-anonymization and aggregation in heterogeneous social
  networks,'' \emph{IEEE Trans. on Dependable and Secure Computing}, vol.~17,
  no.~2, pp. 350--362, 2020.

\bibitem{Feng2011}
J.~{Feng} and A.~K. {Jain}, ``Fingerprint reconstruction: From minutiae to
  phase,'' \emph{IEEE Trans. on Pattern Analysis and Machine Intelligence},
  vol.~33, no.~2, pp. 209--223, 2011.

\bibitem{Cao2015}
K.~{Cao} and A.~K. {Jain}, ``Learning fingerprint reconstruction: From minutiae
  to image,'' \emph{IEEE Trans. on Information Forensics and Security},
  vol.~10, no.~1, pp. 104--117, 2015.

\bibitem{AlRubaie2016}
M.~{Al-Rubaie} and J.~M. {Chang}, ``Reconstruction attacks against mobile-based
  continuous authentication systems in the cloud,'' \emph{IEEE Trans. on
  Information Forensics and Security}, vol.~11, no.~12, pp. 2648--2663, 2016.

\bibitem{Lotte2018}
F.~Lotte, L.~Bougrain, A.~Cichocki, M.~Clerc, M.~Congedo, A.~Rakotomamonjy, and
  F.~Yger, ``A review of classification algorithms for {EEG}-based
  brain-computer interfaces: A 10-year update,'' \emph{Journal of Neural
  Engineering}, vol.~15, 2018.

\bibitem{drwuTLBCI2021}
D.~Wu, Y.~Xu, and B.-L. Lu, ``Transfer learning for {EEG}-based brain-computer
  interfaces: A review of progress made since 2016,'' \emph{IEEE Trans. on
  Cognitive and Developmental Systems}, 2020, in press.

\bibitem{Liang2020}
J.~Liang, D.~Hu, and J.~Feng, ``Do we really need to access the source data?
  {S}ource hypothesis transfer for unsupervised domain adaptation,'' in
  \emph{Proc. 37th Int'l Conf. on Machine Learning}, Vienna, Austria, Jul.
  2020.

\bibitem{Song2017}
C.~Song, T.~Ristenpart, and V.~Shmatikov, ``Machine learning models that
  remember too much,'' in \emph{Proc. ACM SIGSAC Conf. on Computer and
  Communications Security}, Dallas, TX, Oct. 2017, pp. 587--601.

\bibitem{Truex2019}
S.~{Truex}, L.~{Liu}, M.~E. {Gursoy}, L.~{Yu}, and W.~{Wei}, ``Demystifying
  membership inference attacks in machine learning as a service,'' \emph{IEEE
  Trans. on Services Computing}, 2019, in press.

\bibitem{Shokri2017}
R.~{Shokri}, M.~{Stronati}, C.~{Song}, and V.~{Shmatikov}, ``Membership
  inference attacks against machine learning models,'' in \emph{Proc. IEEE
  Symposium on Security and Privacy}, San Jose, CA, May 2017, pp. 3--18.

\bibitem{Fredrikson2015}
M.~Fredrikson, S.~Jha, and T.~Ristenpart, ``Model inversion attacks that
  exploit confidence information and basic countermeasures,'' in \emph{Proc.
  22nd ACM SIGSAC Conf. on Computer and Communications Security}, Denver, CO,
  Oct. 2015, pp. 1322--1333.

\bibitem{drwuCIM2021}
D.~Wu, W.~Fang, Y.~Zhang, L.~Yang, H.~Luo, L.~Ding, X.~Xu, and X.~Yu,
  ``Adversarial attacks and defenses in physiological computing: A systematic
  review,'' \emph{IEEE Computational Intelligence Magazine}, 2021, submitted.

\bibitem{drwuALBCI2019}
X.~Jiang, X.~Zhang, and D.~Wu, ``Active learning for black-box adversarial
  attacks in {EEG}-based brain-computer interfaces,'' in \emph{Proc. IEEE
  Symposium Series on Computational Intelligence}, Xiamen, China, Dec. 2019.

\bibitem{drwuNSR2021}
X.~Zhang, D.~Wu, L.~Ding, H.~Luo, C.-T. Lin, T.-P. Jung, and R.~Chavarriaga,
  ``Tiny noise, big mistakes: Adversarial perturbations induce errors in
  brain-computer interface spellers,'' \emph{National Science Review}, vol.~8,
  no.~4, 2021.

\bibitem{drwuTAR2019}
L.~Meng, C.-T. Lin, T.-P. Jung, and D.~Wu, ``White-box target attack for
  {EEG}-based {BCI} regression problems,'' in \emph{Proc. Int'l Conf. on Neural
  Information Processing}, Sydney, Australia, Dec. 2019.

\bibitem{drwuUAP2021}
\BIBentryALTinterwordspacing
Z.~Liu, L.~Meng, X.~Zhang, W.~Fang, and D.~Wu, ``Universal adversarial
  perturbations for {CNN} classifiers in {EEG}-based {BCIs},'' \emph{Journal of
  Neural Engineering}, vol.~18, no.~4, p. 0460a4, 2021. [Online]. Available:
  \url{https://arxiv.org/abs/1912.01171}
\BIBentrySTDinterwordspacing

\bibitem{Kong2018}
X.~Kong, W.~Kong, Q.~Fan, Q.~Zhao, and A.~Cichocki, ``Task-independent {EEG}
  identification via low-rank matrix decomposition,'' in \emph{IEEE Int'l Conf.
  on Bioinformatics and Biomedicine}, Madrid, Spain, Dec. 2018, pp. 412--419.

\bibitem{Brunner2008}
C.~Brunner, R.~Leeb, G.~M\"{u}ller-Putz, A.~Schl\"{o}gl, and G.~Pfurtscheller,
  ``{BCI} competition 2008 -- {Graz} data set {A},'' \emph{Institute for
  Knowledge Discovery (Laboratory of Brain-Computer Interfaces), Graz
  University of Technology}, vol.~16, pp. 1--6, 2008.

\bibitem{Riccio2013}
A.~Riccio, L.~Simione, F.~Schettini, A.~Pizzimenti, M.~Inghilleri, M.~O.
  Belardinelli, D.~Mattia, and F.~Cincotti, ``Attention and p300-based {BCI}
  performance in people with amyotrophic lateral sclerosis,'' \emph{Frontiers
  in Human Neuroscience}, vol.~7, no.~1, p. 732, 2013.

\bibitem{Zheng2015}
W.-L. Zheng and B.-L. Lu, ``Investigating critical frequency bands and channels
  for {EEG}-based emotion recognition with deep neural networks,'' \emph{IEEE
  Trans. on Autonomous Mental Development}, vol.~7, no.~3, pp. 162--175, 2015.

\bibitem{Hoeller2018}
Y.~H\"{o}ller and A.~Uhl, ``Do {EEG}-biometric templates threaten user
  privacy?'' in \emph{Proc. 6th ACM Workshop on Information Hiding and
  Multimedia Security}, Innsbruck, Austria, Jun. 2018, pp. 31--42.

\bibitem{Yu2020}
K.~Yu, J.~Huang, L.~Tan, G.~Srivastava, X.~Shang, and P.~Chatterjee,
  ``Efficient and privacy-preserving medical research support platform against
  {COVID-19}: A blockchain-based approach,'' \emph{IEEE Consumer Electronics
  Magazine}, vol.~10, no.~2, pp. 111--120, 2020.

\bibitem{Iwendi2020}
C.~Iwendi, S.~A. Moqurrab, A.~Anjum, S.~Khan, S.~Mohan, and G.~Srivastava,
  ``{N-Sanitization}: A semantic privacy-preserving framework for unstructured
  medical datasets,'' \emph{Computer Communications}, vol. 161, no. 2020, pp.
  160--171, 2020.

\bibitem{Bost2015}
R.~Bost, R.~A. Popa, S.~Tu, and S.~Goldwasser, ``Machine learning
  classification over encrypted data,'' in \emph{Proc. Network and Distributed
  System Security Symposium}, San Diego, CA, Feb. 2015.

\bibitem{Dowlin2016}
N.~Dowlin, R.~Gilad-Bachrach, K.~Laine, K.~Lauter, M.~Naehrig, and J.~Wernsing,
  ``{CryptoNets}: Applying neural networks to encrypted data with high
  throughput and accuracy,'' in \emph{Proc. 33rd Int'l Conf. on Int'l Conf. on
  Machine Learning}, New York, NY, Jun. 2016, pp. 201--210.

\bibitem{Kuri2017}
S.~{Kuri}, T.~{Hayashi}, T.~{Omori}, S.~{Ozawa}, Y.~{Aono}, L.~T. {Phong},
  L.~{Wang}, and S.~{Moriai}, ``Privacy preserving extreme learning machine
  using additively homomorphic encryption,'' in \emph{Proc. IEEE Symposium
  Series on Computational Intelligence}, Honolulu, HI, Nov. 2017, pp. 1--8.

\bibitem{Sun2020}
X.~{Sun}, P.~{Zhang}, J.~K. {Liu}, J.~{Yu}, and W.~{Xie}, ``Private machine
  learning classification based on fully homomorphic encryption,'' \emph{IEEE
  Trans. on Emerging Topics in Computing}, vol.~8, no.~2, pp. 352--364, 2020.

\bibitem{Yao1986}
A.~C. {Yao}, ``How to generate and exchange secrets,'' in \emph{Proc. 27th
  Annual Symposium on Foundations of Computer Science}, Toronto, Canada, Oct.
  1986.

\bibitem{Ahlswede1993}
R.~{Ahlswede} and I.~{Csiszar}, ``Common randomness in information theory and
  cryptography. {I}. secret sharing,'' \emph{IEEE Trans. on Information
  Theory}, vol.~39, no.~4, pp. 1121--1132, 1993.

\bibitem{Nikolaenko2013}
V.~{Nikolaenko}, U.~{Weinsberg}, S.~{Ioannidis}, M.~{Joye}, D.~{Boneh}, and
  N.~{Taft}, ``Privacy-preserving ridge regression on hundreds of millions of
  records,'' in \emph{Proc. IEEE Symposium on Security and Privacy}, Berkeley,
  CA, May 2013, pp. 334--348.

\bibitem{Ohrimenko2016}
O.~Ohrimenko, F.~Schuster, C.~Fournet, A.~Mehta, S.~Nowozin, K.~Vaswani, and
  M.~Costa, ``Oblivious multi-party machine learning on trusted processors,''
  in \emph{Proc. 25th USENIX Security Symposium}, Austin, TX, Aug. 2016, pp.
  619--636.

\bibitem{Shaon2017}
F.~Shaon, M.~Kantarcioglu, Z.~Lin, and L.~Khan, ``{SGX-BigMatrix}: {A}
  practical encrypted data analytic framework with trusted processors,'' in
  \emph{Proc. {ACM} {SIGSAC} Conf. on Computer and Communications Security},
  Dallas, TX, Oct. 2017, pp. 1211--1228.

\bibitem{drwuBigData2018}
A.~{Agarwal}, R.~{Dowsley}, N.~D. {McKinney}, D.~{Wu}, C.-T. {Lin}, M.~{De
  Cock}, and A.~C.~A. {Nascimento}, ``Privacy-preserving linear regression for
  brain-computer interface applications,'' in \emph{Proc. IEEE Int'l Conf. on
  Big Data}, Seattle, WA, Dec. 2018, pp. 5277--5278.

\bibitem{Agarwal2019}
A.~{Agarwal}, R.~{Dowsley}, N.~D. {McKinney}, D.~{Wu}, C.-T. {Lin}, M.~{De
  Cock}, and A.~C.~A. {Nascimento}, ``Protecting privacy of users in
  brain-computer interface applications,'' \emph{IEEE Trans. on Neural Systems
  and Rehabilitation Engineering}, vol.~27, no.~8, pp. 1546--1555, 2019.

\bibitem{Zhang2018}
T.~Zhang, Z.~He, and R.~B. Lee, ``Privacy-preserving machine learning through
  data obfuscation,'' \emph{arXiv preprint arXiv:1807.01860}, 2018.

\bibitem{Zhu2017}
T.~{Zhu}, G.~{Li}, W.~{Zhou}, and P.~S. {Yu}, ``Differentially private data
  publishing and analysis: A survey,'' \emph{IEEE Trans. on Knowledge and Data
  Engineering}, vol.~29, no.~8, pp. 1619--1638, 2017.

\bibitem{Dwork2011}
C.~Dwork, ``A firm foundation for private data analysis,'' \emph{Communications
  of the ACM}, vol.~54, no.~1, pp. 86--95, 2011.

\bibitem{McSherry2007}
F.~{McSherry} and K.~{Talwar}, ``Mechanism design via differential privacy,''
  in \emph{Proc. 48th Annual IEEE Symposium on Foundations of Computer
  Science}, Providence, RI, Oct. 2007, pp. 94--103.

\bibitem{Rubinstein2009}
B.~I.~P. Rubinstein, P.~L. Bartlett, L.~Huang, and N.~Taft, ``Learning in a
  large function space: Privacy-preserving mechanisms for {SVM} learning,''
  \emph{Journal of Privacy and Confidentiality}, vol.~4, no.~1, pp. 65--100,
  2012.

\bibitem{Chaudhuri2011}
K.~Chaudhuri, C.~Monteleoni, and A.~D. Sarwate, ``Differentially private
  empirical risk minimization,'' \emph{Journal of Machine Learning Research},
  vol.~12, no.~41, pp. 1069--1109, 2011.

\bibitem{Xiong2020}
X.~Xiong, S.~Liu, D.~Li, Z.~Cai, and X.~Niu, ``A comprehensive survey on local
  differential privacy,'' \emph{Security and Communication Networks}, vol.
  2020, pp. 1--29, 2020.

\bibitem{Shi2011}
E.~Shi, T.~H.~H. Chan, and E.~G. Rieffel, ``Privacy-preserving aggregation of
  time-series data,'' in \emph{Proc. Network and Distributed System Security
  Symposium}, San Diego, CA, Feb. 2011.

\bibitem{Liu2017}
Y.~Liu, W.~Zhang, and N.~Yu, ``Protecting privacy in shared photos via
  adversarial examples based stealth,'' \emph{Security and Communication
  Networks}, vol. 2017, pp. 1--16, 2017.

\bibitem{Pascual2020}
D.~{Pascual}, A.~{Amirshahi}, A.~{Aminifar}, D.~{Atienza}, P.~{Ryvlin}, and
  R.~{Wattenhofer}, ``{EpilepsyGAN}: Synthetic epileptic brain activities with
  privacy preservation,'' \emph{IEEE Trans. on Biomedical Engineering}, 2020,
  in press.

\bibitem{Liu2006}
K.~Liu, H.~{Kargupta}, and J.~{Ryan}, ``Random projection-based multiplicative
  data perturbation for privacy preserving distributed data mining,''
  \emph{IEEE Trans. on Knowledge and Data Engineering}, vol.~18, no.~1, pp.
  92--106, 2006.

\bibitem{Alotaibi2012}
K.~{Alotaibi}, V.~J. {Rayward-Smith}, W.~{Wang}, and B.~{de la Iglesia},
  ``Non-linear dimensionality reduction for privacy-preserving data
  classification,'' in \emph{Proc. Int'l Conf. on Privacy, Security, Risk and
  Trust and Int'l Conf on Social Computing}, Amsterdam, Netherlands, Sep. 2012,
  pp. 694--701.

\bibitem{Niinimaeki2019}
T.~Niinim\"aki, M.~A. Heikkil\"a, A.~Honkela, and S.~Kaski, ``Representation
  transfer for differentially private drug sensitivity prediction,''
  \emph{Bioinformatics}, vol.~35, no.~14, pp. 218--224, 2019.

\bibitem{Denton2015}
E.~L. Denton, S.~Chintala, A.~Szlam, and R.~Fergus, ``Deep generative image
  models using a {Laplacian} pyramid of adversarial networks,'' in \emph{Proc.
  Advances in Neural Information Processing Systems}, Montreal, Canada, Dec.
  2015, pp. 1486--1494.

\bibitem{Kulkarni2018}
V.~Kulkarni, N.~Tagasovska, T.~Vatter, and B.~Garbinato, ``Generative models
  for simulating mobility trajectories,'' \emph{arXiv preprint
  arXiv:1811.12801}, 2018.

\bibitem{SamsonCheung2018}
S.~{Samson Cheung}, H.~{Wildfeuer}, M.~{Nikkhah}, X.~{Zhu}, and W.~{Tan},
  ``Learning sensitive images using generative models,'' in \emph{Proc. 25th
  IEEE Int'l Conf. on Image Processing}, Athens, Greece, Oct. 2018, pp.
  4128--4132.

\bibitem{Acs2019}
G.~{Acs}, L.~{Melis}, C.~{Castelluccia}, and E.~{De Cristofaro},
  ``Differentially private mixture of generative neural networks,'' \emph{IEEE
  Trans. on Knowledge and Data Engineering}, vol.~31, no.~6, pp. 1109--1121,
  2019.

\bibitem{Debie2020}
E.~{Debie}, N.~{Moustafa}, and M.~T. {Whitty}, ``A privacy-preserving
  generative adversarial network method for securing {EEG} brain signals,'' in
  \emph{Proc. Int'l Joint Conf. on Neural Networks}, Glasgow, United Kingdom,
  Jul. 2020, pp. 1--8.

\bibitem{Sion2019}
L.~{Sion}, D.~{Van Landuyt}, K.~{Wuyts}, and W.~{Joosen}, ``Privacy risk
  assessment for data subject-aware threat modeling,'' in \emph{Proc. IEEE
  Security and Privacy Workshops}, San Francisco, CA, May 2019, pp. 64--71.

\bibitem{Orekondy2018}
T.~{Orekondy}, M.~{Fritz}, and B.~{Schiele}, ``Connecting pixels to privacy and
  utility: Automatic redaction of private information in images,'' in
  \emph{Proc. IEEE/CVF Conf. on Computer Vision and Pattern Recognition}, Salt
  Lake City, UT, Jun. 2018, pp. 8466--8475.

\bibitem{Liu2020}
B.~Liu, M.~Ding, S.~Shaham, W.~Rahayu, F.~Farokhi, and Z.~Lin, ``When machine
  learning meets privacy: A survey and outlook,'' \emph{arXiv preprint
  arxiv:2011.11819}, 2020.

\bibitem{Wijesekera2017}
P.~{Wijesekera}, A.~{Baokar}, L.~{Tsai}, J.~{Reardon}, S.~{Egelman},
  D.~{Wagner}, and K.~{Beznosov}, ``The feasibility of dynamically granted
  permissions: Aligning mobile privacy with user preferences,'' in \emph{Proc.
  IEEE Symposium on Security and Privacy}, San Jose, CA, May 2017, pp.
  1077--1093.

\bibitem{Fu2019}
H.~{Fu}, Z.~{Zheng}, S.~{Zhu}, and P.~{Mohapatra}, ``Keeping context in mind:
  Automating mobile app access control with user interface inspection,'' in
  \emph{Proc. IEEE Conf. on Computer Communications}, Paris, France, Apr. 2019,
  pp. 2089--2097.

\bibitem{Robinson2016}
V.~{Robinson} and E.~B. {Varghese}, ``A novel approach for ensuring the privacy
  of {EEG} signals using application-specific feature extraction and {AES}
  algorithm,'' in \emph{Proc. Int'l Conf. on Inventive Computation
  Technologies}, Coimbatore, India, Aug. 2016, pp. 1--6.

\bibitem{Popescu2021}
A.~B. Popescu, L.~A. Taca, C.~I. Nita, A.~Vizitiu, R.~Demeter, C.~Suciu, and
  L.~M. Itu, ``Privacy preserving classification of {EEG} data using machine
  learning and homomorphic encryption,'' \emph{Applied Sciences}, vol.~11,
  no.~16, p. 7360, 2021.

\bibitem{Montenegro2021}
H.~Montenegro, W.~Silva, and J.~S. Cardoso, ``Towards privacy-preserving
  explanations in medical image analysis,'' \emph{arXiv preprint
  arXiv:2107.09652}, 2021.

\bibitem{Schomer2017}
D.~L. Schomer and F.~H.~L. da~Silva, \emph{Niedermeyer's Electroencephalography
  Basic Principles, Clinical Applications, and Related Fields}.\hskip 1em plus
  0.5em minus 0.4em\relax Oxford, UK: Oxford University Press, 2017.

\bibitem{Pan2010}
S.~J. {Pan} and Q.~{Yang}, ``A survey on transfer learning,'' \emph{IEEE Trans.
  on Knowledge and Data Engineering}, vol.~22, no.~10, pp. 1345--1359, 2010.

\bibitem{drwuEA2020}
H.~He and D.~Wu, ``Transfer learning for brain-computer interfaces: A
  {Euclidean} space data alignment approach,'' \emph{{IEEE} Trans. on
  Biomedical Engineering}, vol.~67, no.~2, pp. 399--410, 2020.

\bibitem{Zhang2021}
H.~Zhang, Y.~Zhang, K.~Jia, and L.~Zhang, ``Unsupervised domain adaptation of
  black-box source models,'' \emph{arXiv preprint arXiv:2101.02839}, 2021.

\bibitem{Rahman2020}
S.~A. {Rahman}, H.~{Tout}, H.~{Ould-Slimane}, A.~{Mourad}, C.~{Talhi}, and
  M.~{Guizani}, ``A survey on federated learning: The journey from centralized
  to distributed on-site learning and beyond,'' \emph{IEEE Internet of Things
  Journal}, 2020, i.

\bibitem{Liu2020a}
Y.~{Liu}, Y.~{Kang}, C.~{Xing}, T.~{Chen}, and Q.~{Yang}, ``A secure federated
  transfer learning framework,'' \emph{IEEE Intelligent Systems}, vol.~35,
  no.~4, pp. 70--82, 2020.

\bibitem{Ju2020}
C.~{Ju}, D.~{Gao}, R.~{Mane}, B.~{Tan}, Y.~{Liu}, and C.~{Guan}, ``Federated
  transfer learning for {EEG} signal classification,'' in \emph{Proc. 42nd
  Annual Int'l Conf. of the IEEE Engineering in Medicine and Biology Society},
  Montreal, Canada, Jul. 2020, pp. 3040--3045.

\bibitem{Senanayake2021}
N.~Senanayake, R.~Podschwadt, D.~Takabi, V.~D. Calhoun, and S.~M. Plis,
  ``Neurocrypt: Machine learning over encrypted distributed neuroimaging
  data,'' \emph{Neuroinformatics}, 2021.

\end{thebibliography}

\end{document}